# The Imaging Database for Epilepsy And Surgery (IDEAS)

Peter N. Taylor[1,2,3]*, Yujiang Wang[1,2,3], Callum Simpson[1], Vytene Janiukstyte[1], Jonathan Horsley[1], Karoline Leiberg[1], Beth Little[1,2], Harry Clifford[1], Sophie Adler[4], Sjoerd B. Vos[5,6], Gavin P Winston[3,6], Andrew W McEvoy[3], Anna Miserocchi[3], Jane de Tisi[3], John S Duncan[3]

1. CNNP Lab (www.cnnp-lab.com), Interdisciplinary Computing and Complex BioSystems Group, School of Computing, Newcastle University, Newcastle upon Tyne, United Kingdom

2. Faculty of Medical Sciences, Newcastle University, Newcastle upon Tyne, United Kingdom

3. UCL Queen Square Institute of Neurology, Queen Square, London, United Kingdom

4. UCL Great Ormond Street Institute of Child Health, London, United Kingdom

5. Centre for Medical Image Computing, Department of Computer Science, UCL, London, United Kingdom

6. Centre for Microscopy, Characterisation, and Analysis, The University of Western Australia, Nedlands, Australia

7. Department of Medicine, Division of Neurology, Queen's University, Kingston, Canada

* peter.taylor@newcastle.ac.uk

## Abstract

Magnetic resonance imaging (MRI) is a crucial tool to identify brain abnormalities in a wide range of neurological disorders. In focal epilepsy MRI is used to identify structural cerebral abnormalities. For covert lesions, machine learning and artificial intelligence algorithms may improve lesion detection if abnormalities are not evident on visual inspection. The success of this approach depends on the volume and quality of training data.

Herein, we release an open-source dataset of preprocessed MRI scans from 442 individuals with drug-refractory focal epilepsy who had neurosurgical resections, and detailed demographic information. The MRI scan data includes the preoperative 3D T1 and where available 3D FLAIR, as well as a manually inspected complete surface reconstruction and volumetric parcellations. Demographic information includes age, sex, age of onset of epilepsy, location of surgery, histopathology of resected specimen, occurrence and frequency of focal seizures with and without impairment of awareness, focal to bilateral tonic-clonic seizures, number of anti-seizure medications (ASMs) at time of surgery, and a total of 1764 patient years of post-surgical follow up. Crucially, we also include resection masks delineated from post-surgical imaging.

To demonstrate the veracity of our data, we successfully replicated previous studies showing long-term outcomes of seizure freedom in the range of around 50%. Our imaging data replicates findings of group level atrophy in patients compared to controls. Resection locations in the cohort were predominantly in the temporal and frontal lobes.

We envisage our dataset, shared openly with the community, will catalyse the development and application of computational methods in clinical neurology.

# Introduction

Large-scale sharing of raw MRI scan data is commonplace in the neurosciences[1], and has led to substantial advances in our understanding of brain function and dysfunction[2]. Such advances are enabled by the association of high quality clinical and demographic metadata with the scan. In neurological conditions such as Alzheimer's[3], autism[4], ADHD[5], Parkinson's[6], and traumatic brain injury[7] there are large MRI data sets available for research. In epilepsy, data sharing was pioneered in the field of seizure prediction, with annotated EEG data available for hundreds of patients[8–10]. However, given that MRI is crucial in the clinical management of epilepsy, it is surprising that relatively little data are openly available, particularly with high quality clinical and demographic information, but note [11–15,45] for related work.

In this study we share anonymised MRI scans from the National Hospital for Neurology and Neurosurgery, UCLH, acquired in 442 individuals with drug-refractory focal epilepsy who proceeded to neurosurgical resection. We also share anonymised demographic and clinical information for all subjects, and masks of subsequently resected tissue. To verify these data we replicate two previous landmark studies[14,25].

# Methods

## Study approval

This study of anonymised data that had been previously acquired was approved by the Health Research Authority, without the necessity to obtain individual subject consent (UCLH epilepsy surgery database: 22/SC/0016), and by the Database Local Data Monitoring Committee. Individuals who declined for their data to be used in anonymised research were not included in the research database.

## Patient and data selection

We identified all eligible people who had neurosurgical resections for drug-resistant focal epilepsy at the National Hospital for Neurology & Neurosurgery, London, UK between 01 January 2003 and 30 June 2022 (n=625). From these we excluded individuals with previous neurosurgery (n=38), and those who had not had pre-operative T1 weighted MRI scans of sufficient quality at the Chalfont Centre for Epilepsy (n=145). Sufficient MRI quality was determined based on visual inspection by either PNT, BL, or VJ and considered aspects including motion artefact, ringing, and field of view completeness. Our final dataset contains pre-operative MRI for 442 individuals, and a resection mask for 433 of those individuals, along with most demographic and clinical data.

## Clinical & demographic information

We extracted the following information from the UCLH epilepsy surgery database, that is prospectively and regularly updated. Data included: age at MRI scan, sex, age at surgery, type and location of surgery, post-operative pathology, age of epilepsy onset, occurrence and frequency of focal seizures with and without impairment of awareness, focal to bilateral tonic-clonic seizures, number of antiseizure medications (ASM) at time of surgery, yearly outcomes of seizure freedom assessed using the ILAE classification[39].

## MRI scan information

All pre-operative 3D-T1-weighted scans were acquired using one of two 3T GE scanners. All subjects included in the release have T1-weighted MRI included, whilst 420 subjects also had FLAIR scans acquired in the same session included. Common acquisition protocols used include 3D fast spoiled gradient echo (FSPGR) at a resolution of 0.9375 x 0.9375 x 1.1mm, or magnetisation-prepared rapid acquisition gradient echo (MPRAGE) with a resolution of 1 x 1 x 1mm. Detailed acquisition parameters for each individual scan are shared as part of the data release in accompanying JavaScript object notation (JSON) format using the brain imaging data structure (BIDS).

## Quality control

Pre-operative T1 weighted scans were processed with the FreeSurfer 7.3.2 software pipeline 'recon-all'[16]. The pipeline performed segmentation and parcellation of cortical tissues. Outputs were then visually inspected following established quality control protocols[14]. We paid particular attention to pial and white matter surfaces. Following visual inspection, some reconstructed surfaces were deemed insufficient and manual edits were made using control points and dura edits to improve their accuracy. For some subjects, particularly those with gross pathology, we noted that surface reconstruction may still be imperfect.

Once a scan and segmentation of sufficient quality was identified, we exported region volumes for deep brain areas, and additional thicknesses and surface areas for neocortical regions using the FreeSurfer command 'aparcstats2table'. These volume, thickness and surface area tables are shared as part of our data release.

## Data harmonisation and normative modelling

Cortical thickness and volume measurements depend on a subject's age, sex, and the scanning protocol used for data acquisition. It is therefore necessary to account for these covariates to investigate the effects of epilepsy. For this, we used a combination of two methods[46,47]. First, a data harmonisation technique, ComBat, which removes scanner differences whilst accounting for and preserving covariate and pathology effects. Second, a normative model, which describes population-level trends in the relationship between variables. Specifically we modelled age, sex, and cortical thickness or volume, and estimated the expected variance in these trends to derive subject-specific deviations[48].

Both ComBat and normative modelling require an inference of age and sex effects of the healthy population. We therefore used data from two large public healthy control datasets, NKI (n=833) and OASIS (n=542)[18,19], for model training. This was in addition to 100 healthy control scans acquired on the same scanners as the patient cohort, which informed on effects of scanning protocols and served as a reference cohort for subsequent abnormality calculations.

For data harmonisation with ComBat, we clustered subjects based on the similarity of their scanning parameters. ComBat then modelled effects of age and sex, as well as the offset and variance of each cluster, so that covariate effects could be retained in the data whilst effects of scanning acquisition were removed[17].

After data harmonisation, we fitted a generalised additive model (GAM) in each region to the healthy controls. For cortical thickness, we used the model formula

$$t \sim 1 + s(age) + sex,$$

where *t* is thickness, *s(age)* is a smooth function of age, and *sex* is a fixed effect.
For subcortical volume, we used the model

$$v \sim 1 + s(age) + sex + ICV,$$

where *v* is volume, *s(age)* is a smooth function of age, and *sex* and *ICV* (intracranial volume) are fixed effects. To apply this normative model, we predicted values of thickness/volume for each patient and the 100 healthy controls which were acquired on the same scanners. We calculated residuals, i.e. differences between the observed and predicted values, which removed the age and sex effects from the data.

## Abnormality calculation

For each subject, we calculated regional abnormalities in cortical thickness and subcortical volume. This was done by z-scoring the residuals, using the mean and standard deviation in each region across the residuals of the 100 reference controls:

$$Z_{ij} = \frac{R_{ij} - \mu_i}{\sigma_i},$$

where $Z_{ij}$ is the z-scored abnormality in region $i$ for subject $j$, $R$ is the residual, $\mu$ is the mean residual across controls and $\sigma$ is the standard deviation of the residuals across controls. These abnormalities quantify how many standard deviations a subject's regional thickness or volume is away from the control population.

Similar abnormalities were also calculated for healthy controls. In this scenario, each control was held out from the control cohort before the calculation of $\mu$ and $\sigma$ used for its z-scoring.

For patients with mesial temporal lobe epilepsy (mTLE), we quantified the extent of alterations in cortical thickness or subcortical volume across the cohort using Cohen's d. In each region, we calculated the effect size of the difference in abnormality between the cohort of patients with mTLE and healthy controls, and plotted the results using the ENIGMA toolbox[20]. This approach is broadly similar to that described previously[14].

### Generation of resection masks

We used postoperative imaging, in the 433 individuals for whom these were available, to generate masks of the tissue that was subsequently resected. Masks were initially generated automatically using a custom-built software pipeline before manual edits of pipeline outputs were performed where needed. The mask generation pipeline comprised three steps. First, data were processed through FastSurfer[21] to label brain regions and the lobe where the resection had taken place was identified. Second, the ANTs registration tool[22] was used to perform a two-part alignment of the pre-operative and post-operative images to compensate for any distortion of remaining brain tissue into the resection cavity seen in the post-operative image. The registration used the lobe information from step 1 to improve registration accuracy. The third step used a classification algorithm ATROPOS[23] to find differences between the registered images. Step three was then followed by manual checks and edits, when needed, to ensure that the produced resection mask was within and accurately filled anatomical boundaries[24].

## Data de-identification and anonymisation

MRI scans were converted to NIFTI format to remove any identifying information in scan headers. Individual faces were removed from all MR images by multiplying a dilated binarised brain mask with the original image. All scans were then visually inspected to ensure that the nose, mouth, and ears were not visible.

Date of birth and date of scan or surgery are not shared publicly since these are protected as potentially identifiable information. Instead, data are shared in categories (e.g. 18-22, 23-27, 28-32 years old). Annual ILAE surgery outcomes of up to five years are shared when available. Data beyond five years are not shared to maintain anonymity regarding date of surgery.

## Data availability

Data will be shared on the openneuro.org platform and publicly searchable without restriction upon acceptance of the manuscript. For review purposes, links to data can be found in Table S3.

# Results

## The IDEAS dataset

Demographic information was available for all patients included in our final data release (n=442). Post-operative imaging to delineate a resection mask was also available for most subjects (N=433, 98%). Table 1 summarises the shared data which include: sex, history of FBTCS (12 months pre-surgery), history and frequency of focal seizures with impaired awareness (12 months pre-surgery), history of status epilepticus, histopathological findings, clinical MRI findings, number of anti-seizure medications at time of surgery, age at MRI scan. Figure 1 presents data from a typical case.

Table 1. Summary of available data

|  | Number of subjects with data shared | Notes |
|---|---|---|
| T1w MRI | 442 |  |
| FLAIR MRI | 411 |  |
| Resection mask | 433 |  |
| Sex | 446 |  |
| Age of epilepsy onset | 442 |  |
| History of FUS | 446 | N=409=true |
| Frequency of FUS | 406 of 409 |  |
| History of FBTCS | 442 | N=343=true |
| Frequency of FBTCS | 330 of 343 |  |
| History of SE | 442 | N=53=true |
| Side of resection | 442 |  |
| Resection type | 442 |  |
| Pathology | 442 |  |
| Number of ASMs | 442 |  |
| Age of epilepsy surgery | 442 |  |
| 12 month ILAE outcome | 427 |  |
| 24 month ILAE outcome | 403 |  |
| 36 month ILAE outcome | 356 |  |
| 48 month ILAE outcome | 311 |  |
| 60 month ILAE outcome | 267 |  |

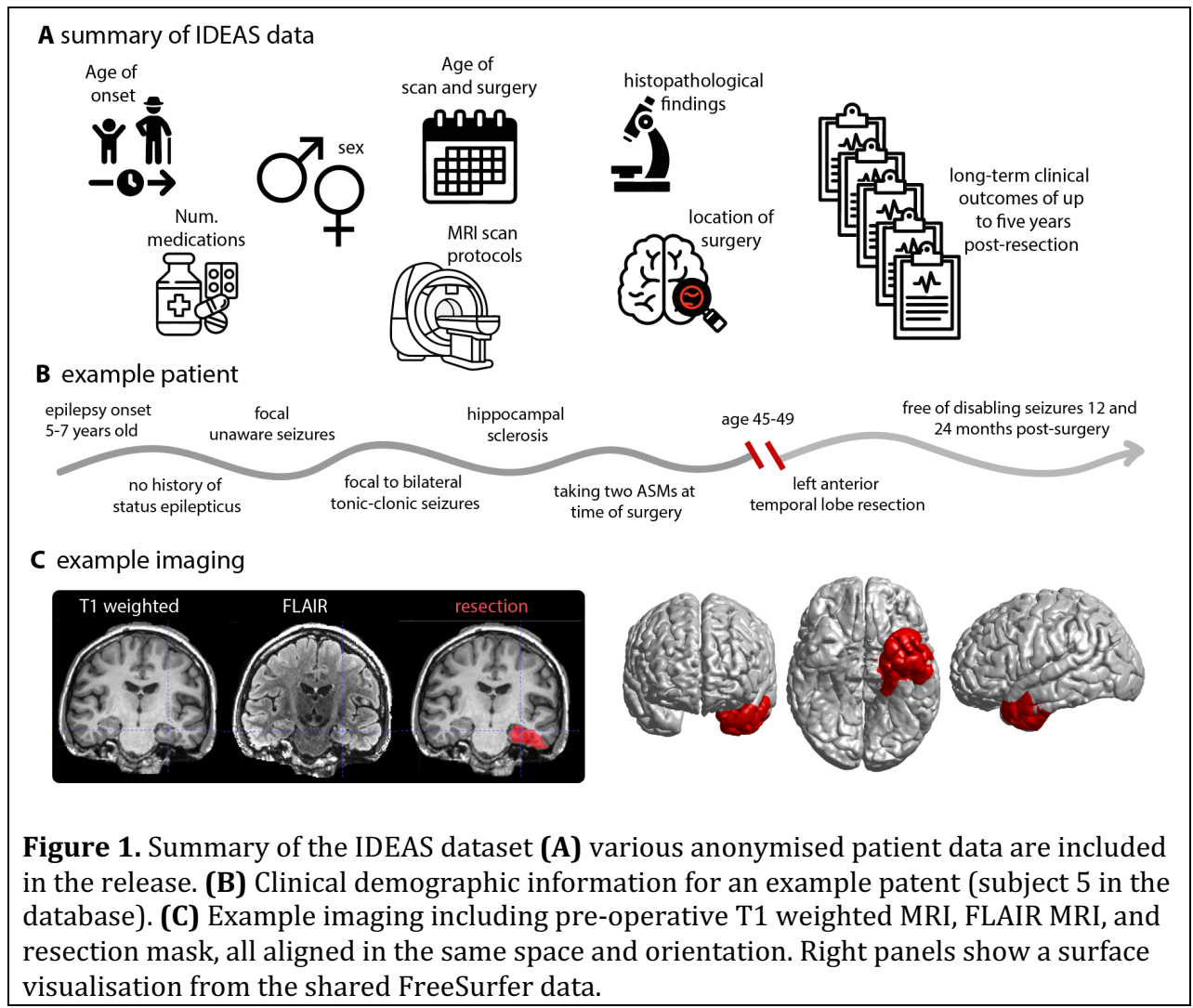

**Figure 1.** Summary of the IDEAS dataset **(A)** various anonymised patient data are included in the release. **(B)** Clinical demographic information for an example patent (subject 5 in the database). **(C)** Example imaging including pre-operative T1 weighted MRI, FLAIR MRI, and resection mask, all aligned in the same space and orientation. Right panels show a surface visualisation from the shared FreeSurfer data.

## Five-year outcome of adult epilepsy surgery, patterns of seizure remission, and relapse

For clinical metadata, we replicate a previous study demonstrating long-term outcomes from epilepsy surgery[25]. Post-operative 12-month outcomes of seizure-freedom were available for 422 patients. Five subjects had multilobar resections. At 12 months, 243 (58%) of the remaining 422 patients were completely seizure-free (figure 2A). After five years of follow up, patients with extra-temporal lobe epilepsy surgery had the lowest estimated proportion of continuous seizure-freedom (44% and 42% for lesionectomy and lobar resection respectively).

Patients with focal aware seizures (FAS) in the first 24 months after surgery were significantly more likely to relapse to have seizures with impaired awareness in the subsequent years (figure 2B, log rank test p<0.001). Sample sizes at each year of follow-up are presented in supplementary table S2. These findings align with those described previously[25].

There were large ranges of age of epilepsy onset (median 13, IQR 14 years), age at surgery (median 37, IQR 17 years) and epilepsy duration (mean 21, IQR 20 years). The minimum age at surgery was 16 years old, with the majority over 18, reflecting our clinical practice as an adult epilepsy centre. None of these variables were significantly related to post-operative seizure outcomes (figure 2C,D,E).

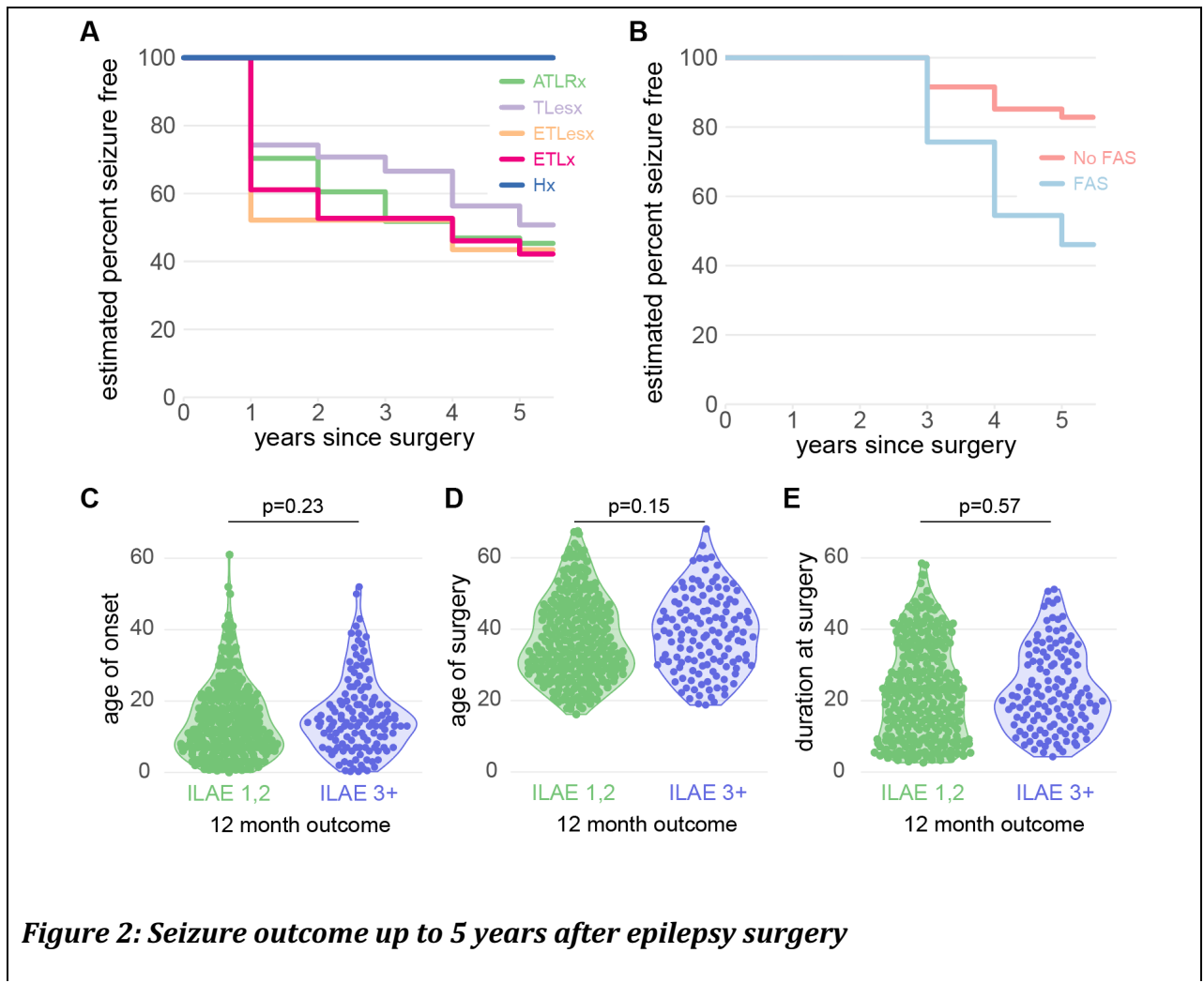

*Figure 2: Seizure outcome up to 5 years after epilepsy surgery*

> ***A*** *Survival plot of proportion of patients remaining seizure-free following epilepsy surgery and with available follow-up at yearly intervals, showing time to first seizure. Each coloured line represents a different surgical procedure (ATLRx=anterior temporal resection. TLesx=temporal lesionectomy. ETLesx=extratemporal lesionectomy. ETLx=extratemporal resection. Hx=hemispherectomy).*
>
> ***B*** *Survival plot of proportion of patients who did (blue), and did not (red), have focal aware seizures (FAS) following surgery in years 1,2, remaining free of seizures with impaired awareness at subsequent years.*
>
> ***C,D,E:*** *12-month seizure outcome did not differ by age of onset of epilepsy, (C), age at surgery (D), or duration of epilepsy (E). All results broadly replicate those described previously in more extensive cohorts from 1990-2012[25,40].*

## Structural brain abnormalities in focal epilepsies

We investigated brain alterations in mTLE, using similar methods to those described previously[14]. Figure 3 shows effect size differences (Cohen's d) in cortical and subcortical brain regions for left, and right, mesial temporal lobe epilepsy (mTLE) (N=122 and N=85 respectively), compared to healthy controls (N=100).

Reduced volumes were noted in several regions including the ipsilateral hippocampus and thalamus. Neocortical thickness was also reduced in several areas. Specifically, in left mTLE the cortical thickness of the caudal middle frontal gyrus was the most reduced ipsilaterally (d=-0.61), followed by the inferior parietal cortex (d=-0.58) and the contralateral precentral gyrus (d=-0.58). In right mTLE, neocortical volumes were reduced in the parietal lobe. Full effect sizes for all regions in all individuals are included in the data release tables.

Other imaging features available in the IDEAS database are available as direct outputs from the FreeSurfer recon-all pipeline, along with fully processed FreeSurfer volumes and surfaces. Outputs include neocortical volumes, surface area and thickness for a further seven parcellations (multi-scale Lausanne[41], Destrieux[42], HCP-MMP1[43], and Whittaker[44]). In addition, volumes of brainstem subregions, hippocampal and amygdala subunits, and hypothalamus segmentation are also available.

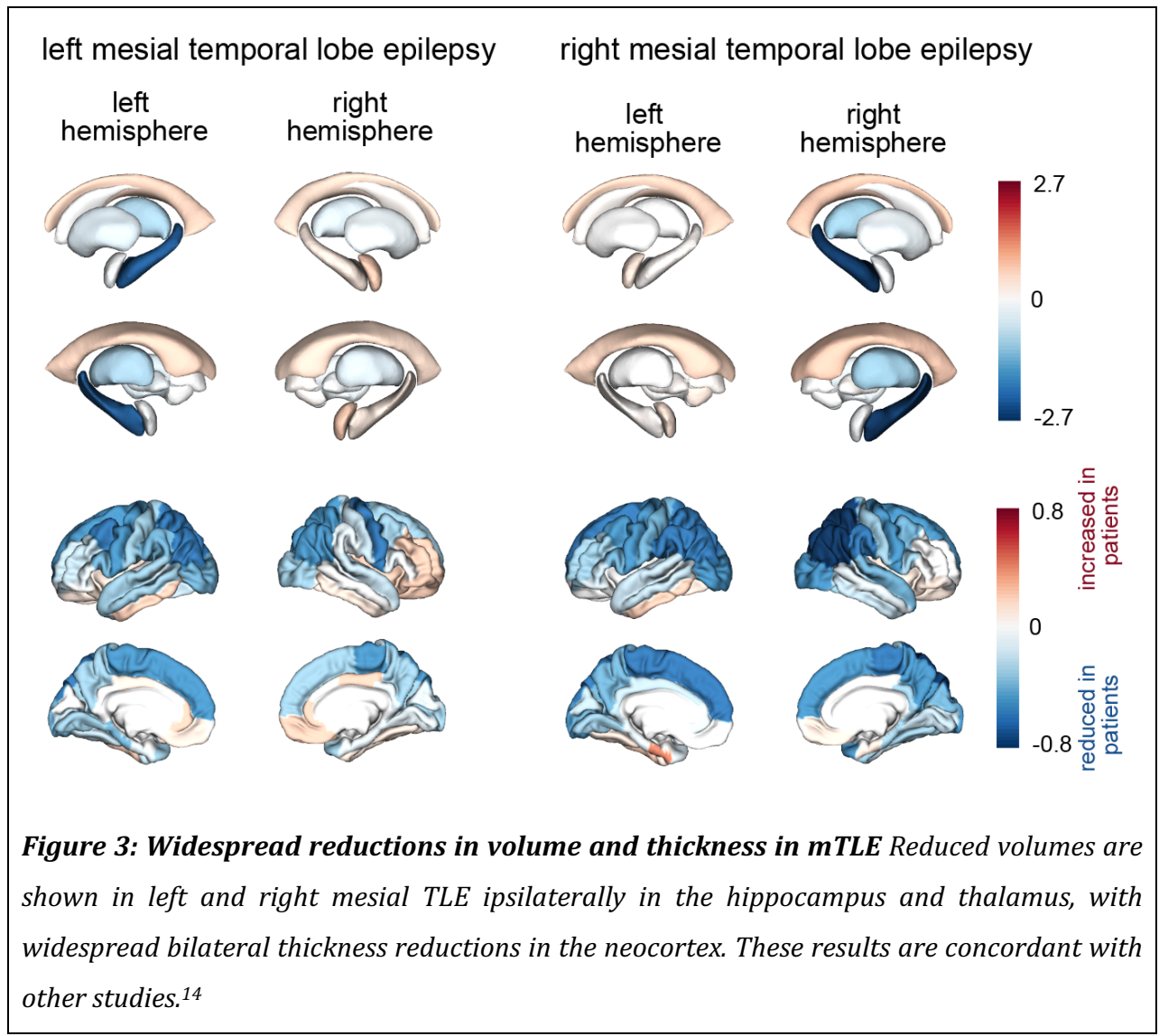

***Figure 3: Widespread reductions in volume and thickness in mTLE*** *Reduced volumes are shown in left and right mesial TLE ipsilaterally in the hippocampus and thalamus, with widespread bilateral thickness reductions in the neocortex. These results are concordant with other studies.*[14]

## Resection masks & locations

Resection masks were generated for N=433 subjects including N=356 individuals with TLE, and N=58 with frontal lobe resections. Masks are visualised in figure 4, overlaid in the same space. Individual resection masks are available within the IDEAS database.

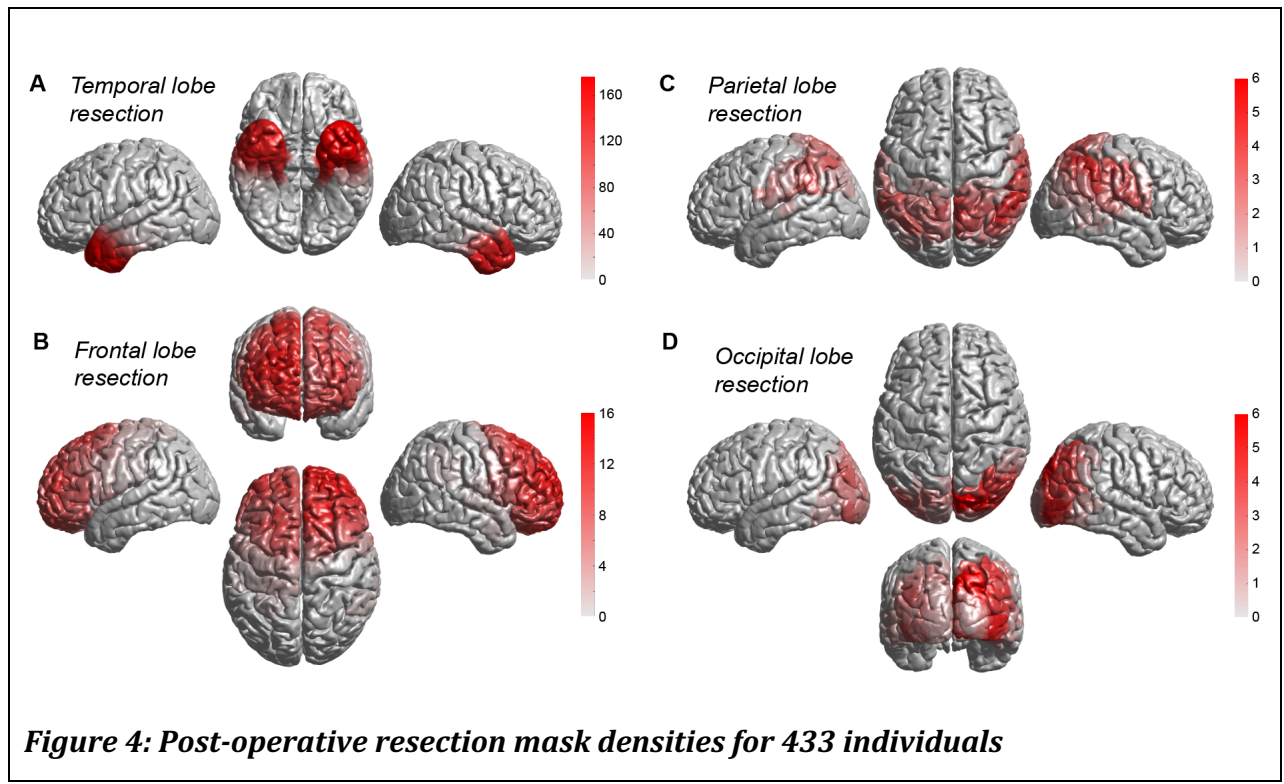

*Figure 4: Post-operative resection mask densities for 433 individuals*

## Discussion

Here we present the Imaging Database for Epilepsy And Surgery (IDEAS). The database contains a large, anonymised sample of neuroimaging, and clinical metadata. The data are organised in an easy-to-use manner, and amenable to future research. Data have been pre-processed, checked for quality and organised using BIDS format. We present demonstrative use-cases for the data, replicating several recent key publications. Data are openly available for the research community.

With the advent of advanced machine learning models and artificial intelligence, we are experiencing a paradigm shift in radiology and beyond. In some task-specific scenarios, AI already exceeds human performance[26]. In epilepsy, recent work has applied AI techniques for lateralisation[31–33] and localisation[27–30] of epileptogenic tissues. Replication of these algorithms on external datasets is crucial to achieve clinical translation. Our data release, including gold standard resections and patient outcome measures represent an opportunity for the field to progress in this area by providing real-world training and test data.

To gain a mechanistic understanding of epilepsy, high quality clinical and demographic data are required. A key component of our data release is the richness of the metadata. Several studies previously reported clinical and demographic factors to be related to post-operative seizure outcomes[34,35]. For example, Jehi et al (2015)[34] reported an association with epilepsy duration, whilst others[35, 36] report FBTCS, being associated with reduced chance of a seizure free outcome. We also share histopathology classification for the resected tissue, which we envisage may be of use to derive MRI markers of specific pathology[37]. These collated clinical and demographic features, in a cohort of this size, are rare in epilepsy. Their associations will be vital to investigate underlying mechanisms.

A crucial facet of our data is that the imaging and metadata are linked. The linking of these data is vital to enable research into our understanding of the pathophysiology and functional anatomy of epilepsy, particularly at the level of the individual. Until now such data on a large scale was unavailable in epilepsy. To facilitate scientific endeavours we have shared all imaging data in a preprocessed format of z-scores, in addition to the raw MRI data. We have also included the metadata in easy-to-read spreadsheets to facilitate future research.

Data quality is crucial. We therefore performed multiple rounds of quality control. For T1-weighted MRI data, all scans were visually inspected, and manual edits made to preprocessed data outputs from FreeSurfer when necessary. Clinical and demographic data, were entered prospectively, verified, and regularly updated, since the inception of the UCLH epilepsy surgery database in 1990. All code to process and analyse the data was verified by multiple people. We are sharing the data on the OpenNeuro platform[1], which is version controlled, and we will provide updates when necessary. We welcome any feedback from the community regarding data issues and commit to continue supporting its continued use.

A further key feature of these data is the inclusion of masks delineating the resection zone. In individuals who are seizure-free in the long-term after epilepsy surgery it can be assumed that at least part of the epileptogenic zone (EZ) is located within the resection zone. It cannot be assumed that there is a one-to-one correspondence between the resection and EZ in seizure-free patients. The resection may be larger than the EZ, for example when surgical access was required via non-epileptogenic tissue. Furthermore, the resection may be smaller than the EZ in seizure-free patients, if the patient is still on anti-seizure medication. Thus, the resection masks represent a silver standard approximation of the EZ, and should be regarded as such. Nonetheless, such a silver standard may still be sufficient for training algorithms for prediction[38].

Full anonymisation of the data that are shared was an important consideration and we took several steps to protect against identification of individuals. For MRI data, these steps included removal of header information from imaging files using NIFTI format and removal of potentially identifying facial features from MRI. Clinical and demographic data were extracted from the clinical database by the clinical team and these and MRI data were pseudo-anonymised prior to any data processing and linking of MRI and clinical data. All data were irrevocably anonymised prior to preparation for being shareable. Important steps to ensure anonymisation included limiting disclosed follow-up to 5 years from surgery, and reducing specificity of age at surgery, age of onset and duration of epilepsy to 5 year epochs.

The use of previously acquired clinical and investigatory data for research, without the requirement for individual patient consent was, approved by Information Governance of UCL Hospitals, the UCLH/UCL Joint Research Office, the Health Research Authority (UCLH epilepsy surgery database: 22/SC/0016) and the Database Data Monitoring Committee. A key requirement is that all data shared externally to UCL and UCLH are totally and irrevocably anonymised.

Although large and richly characterised, our data have limitations. First, our data reflects resective adult epilepsy surgery clinical practice at a single centre over the last 20 years. As such, no individuals who received laser interstitial thermal therapy or thermocoagulation are included. Second, our surgical practice is confined to adults, and any inferences with regard to paediatric epilepsy are limited. Third, our data release of 3D-T1-weighted and T2-FLAIR MRI represents only part of the information used during pre-surgical evaluation. Many of the individuals having neurosurgical treatment of drug-refractory focal epilepsy also underwent acquisition of data from other modalities, including diffusion MRI, PET, SPECT high density scalp EEG, MEG and intracranial EEG. Neuropsychological and neuropsychiatric data are also typically considered during presurgical evaluation. We hope to be able to include these data in the coming years.

We hope that the IDEAS data will be a valuable resource for the epilepsy research community that will catalyse efforts in data science research in epilepsy. We welcome contributions from other sites, and propose our metadata headers as common data elements to be used for cross site consistency. We will be pleased to concatenate contributions from other centres with our own to establish a larger, multi-centre resource.

# Acknowledgements

We thank members of the Computational Neurology, Neuroscience & Psychiatry Lab (www.cnnp-lab.com) for discussions on the analysis and manuscript. We are grateful to the UCLH Epilepsy Surgery Database local Data Monitoring Committee. P.N.T. and Y.W. are both supported by UKRI Future Leaders Fellowships (MR/T04294X/1, MR/V026569/1). Y.W. and B.L. are supported by EPSRC (EP/Y016009/1). G.P.W. and acquisition of control data was supported by the MRC (G0802012, MR/M00841X/1). JSD, JdT are supported by the NIHR UCLH/UCL Biomedical Research Centre. This work was supported by Epilepsy Research UK (grant number P1904).